# STEGANALYSIS USING COLOUR MODEL CONVERSION

P.Thiyagarajan[#1], G.Aghila[#2] and V. Prasanna Venkatesan[#3]

[#]CDBR-SSE Lab Department of Computer Science, Pondicherry University, Puducherry 605 014
[1]thiyagu.phd@gmail.com, [2]aghilaa@gmail.com,
[3]prasanna_v@yahoo.com

## ABSTRACT

*The major threat in cyber crime for digital forensic examiner is to identify, analyze and interpret the concealed information inside digital medium such as image, audio and video. There are strong indications that hiding information inside digital medium has been used for planning criminal activities. In this way, it is important to develop a steganalysis technique which detects the existence of hidden messages inside digital medium. This paper focuses on universal image steganalysis method which uses RGB to HSI colour model conversion. Any Universal Steganalysis algorithm developed should be tested with various stego-images to prove its efficiency. The developed Universal Steganalysis algorithm is tested in stego-image database which is obtained by implementing various RGB Least Significant Bit Steganographic algorithms. Though there are many stego-image sources available on the internet it lacks in the information such as how many rows has been infected by the steganography algorithms, how many bits have been modified and which channel has been affected. These parameters are important for Steganalysis algorithms and it helps to rate its efficiency. Proposed Steganalysis using Colour Model has been tested with our Image Database and the results were affirmative.*

## 1. INTRODUCTION

Steganography is the practice of concealing the very presence of message during communication [1]. Like two sides of the coin, Steganography has both advantages and disadvantages. It depends on the person how he uses it for example if it is in the hands of the scientist he may use it for the military purpose or if it is in the hands of the terrorists he may make use of steganography to snatch the attack plan among his team members and communicate via internet. In the later case it is more important to detect the image which is used by criminals. The technique which is used to identify the images which is hands of terrorist that contains the secret message is called as Steganalysis. Our work aims to develop Universal Steganalysis algorithm and to test the proposed Steganalysis algorithm on variety of stego-images. Generally Steganalysis algorithm uses Statistical methods to classify stego-image from cover Image but our proposed method uses colour model conversion and visual perception to differentiate Stego and cover image .Section 2 gives outline about Steganography, Section 3 gives brief outline about Steganalysis, Section 4 outlines about various Steganalysis algorithm and various Image Steganography Tools and emphasis the need for Universal Steganalysis algorithm, Section 5 describes the Architecture of the proposed Universal Steganalysis algorithm tested on images obtained from our own Stego-Image Generator (SIG) Tool, Section 6 shows the experimental results of the proposed method and Section 7 concludes the paper.

## 2. STEGANOGRAPHY

Steganography can be best explained through prisoner's problem Alice wishes to send a secret message to Bob by hiding information in a clean image. The stego image (clean image + secret message) passes through Wendy (a warden) who inspects it to determine if there is anything suspicious about it. Wendy could perform one or several scan to decide if the message from Alice to Bob contains any secret information. If the decision is negative then Wendy forwards the message to Bob in that case Wendy acts as a passive warden. On the other hand, Wendy can take a conservative approach and modify all the messages from Alice to Bob irrespective of whether any information is hidden by Alice or not.

In this case, Wendy is called an active warden. Of course, Wendy will have constraints such as the maximum allowable distortion when modifying the message etc. For example, if the clean messages are digital images, then Wendy cannot modify the stego message to an extent that perceptually significant distortions are induced. Fig 1 represents the pictorial representation of how message is embedded and extracted using steganography algorithm.

LSB methods are most commonly used steganography techniques. Least Significant Bit of some or all of the bytes inside an image is changed to a bit of the secret message. When using a 24-bit RGB Color image, a bit of each of the red, green and blue color components can be used. In other words, one pixel can store 3 bits of secret message.

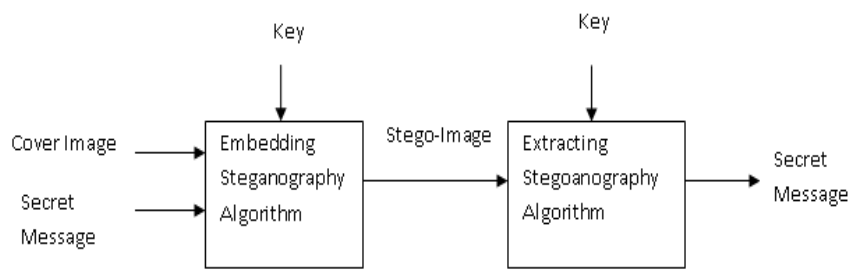

Fig. 1 Steganography Process Diagram

Components

- Secret Message - The message to be embedded
- Cover Image – An image in which Secret Message will be embedded.
- Stego Image - Cover image that contain embedded message.
- Key – Additional data that is needed for embedding and extracting process.
- Embedding Steganography Algorithm - Steganography Algorithm used to embed secret message with cover image.
- Extracting Steganography Algorithm - Inverse function of embedding, in which it is used to extract the embedded message (secret message) from stego image.

## 3. STEGANALYSIS

Steganography can be applied in numerous fields such as authentication, secret communication in military, banking etc., however it also depends on the person who is using. There are strong indications that steganography has been used for planning criminal activities [2]. In this way, it is important to detect the existence of hidden messages in digital files. As with cryptography and cryptanalysis, Steganalysis is defined as the art and science of breaking the security of steganography systems. The goal of steganography is to conceal the existence of a secret message. While the goal of Steganalysis is to detect that a certain file contains embedded data. The stegosystem can be extended to include scenarios for different attacks [4] similar to the attacks on cryptographic systems. In general, extraction of the secret message could be a harder problem than mere detection. The challenges of steganalysis are listed below

a) To get Stego-Image Database
b) To develop Universal or Specific Steganalysis Algorithm
c) To test the Steganalysis algorithm against different payload Stego-Images to check its robustness
d) To test the Steganalysis algorithm from various categories of images such as animals, fruits, natural scene etc.,
e) Detection of presence of hidden message in cover signal
f) Identification of embedding algorithm
g) Estimation of embedded message length
h) Prediction of location of hidden message bits
i) Estimation of secret key used in the embedding algorithm
j) Estimation of parameter of embedding algorithm
k) Extraction of hidden message

This paper mainly focuses on the issues pertaining to creation of Stego-Image database and Steganalysis methods which falls on the first five issues listed above.

## 4 . LITERATURE REVIEW

Detailed literature survey on Steganography Image tool and Steganalysis algorithm related to pixel colour values has surveyed and it is discussed in this section.

### 4.1 Stego-Image Tool

We have surveyed 15 Stego-images generation tools. Detailed survey [5] of the below tools have been made and they have been classified according to the format of stego-image it produces, availability of code. Table 1 gives details about the various stego-image generation tools where the code is available [6] [7] [8] [9] [10] [11]. Table 2 gives the detail about the Stego-image generation tool where the code is not available but its executable is available [12] [13] [14] [15] [16] [17].

Table 1: Image Steganography Tools with Source Code

| S.No | Name of Image Steganography Tool | Format | Availability of Code |
|------|----------------------------------|--------|----------------------|
| 1 | Blind slide | BMP | Yes |
| 2 | Camera Shy | JPEG | Yes |
| 3 | Hide4PGP | BMP | Yes |
| 4 | JP Hide and Seek | JPEG | Yes |

| 5 | Jsteg Jpeg | JPEG | Yes |
| 6 | Mandelsteg | GIF | Yes |
| 7 | Steghide | BMP | Yes |
| 8 | wbStego | BMP | Yes |

Table 2: Image Steganography Tools without Source Code

| S.No | Name of Image Steganography Tool | Format | Availability of Code |
| --- | --- | --- | --- |
| 1 | Camouflage | PNG | No |
| 2 | Hide & Seek | BMP,GIF | No |
| 3 | S-Tools | BMP | No |
| 4 | Steganos | BMP | No |
| 5 | StegMark | BMP,GIF,PNG | No |
| 6 | Invisible Secrets | BMP,JPEG,GIF | No |
| 7 | Info Stego | BMP,JPEG,PNG | No |

### 4.2 Steganalysis Method

There are many steganalysis methods reported in the literature survey. Some of the steganalysis methods based on the colour of the pixel are discussed here. Pfitzman and Westfeld [18] introduced a powerful statistical attack that can be applied to any steganography technique in which a set of Pairs of Values (PoVs) are used to detect the presence of secret message. Pfitzman et.al exploited the fact that any Steganographic techniques change the frequency of pair of value during message embedding process. This method was effective in detecting Stego-images generated from variety of Steganography algorithms.

Mitra et.al. [19] described a detection theory based on statistical analysis of pixel pairs using their RGB components to detect the presence of hidden messages in LSB steganography. Here the Cover image and Stego-Image is differentiated by threshold value. This threshold value depends on the correlation between pixel pairs in terms of colour components. Apart from threshold value unique colour component also plays a major role in finding out the Stego-Images. Raja et.al. [20] explains LSB Steganalysis method through Colour Pair Value Variable threshold which is derived from Colour Density of the image.

Parametric statistical steganalysis approaches [2, 24, 25, 26] assume certain parametric statistical model for the cover image, stego image and the hidden message. Steganalysis is formulated as a hypothesis testing problem, namely, H0: no message (null hypothesis) and H1: message present (alternate hypothesis). A statistical detection algorithm is then designed to test between the two hypotheses.

## 5. ARCHITECTURE OF PROPOSED STEGANALYSIS METHOD

The architecture of the proposed method is shown in Fig 2.The proposed architecture can be divided in to two major components. They are

a) Stego-Image Generator (SIG) Tool

b)   Steganalysis using HSI Colour Model

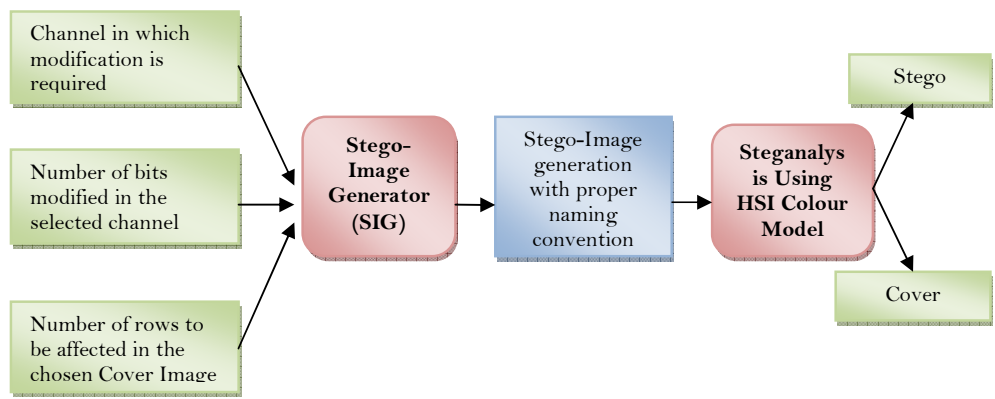

Fig. 2 Architecture of Steganalysis Method using HSI Colour Model along with Stego-Image Generator (SIG) Tool

## 5.1 Stego-Image Generator (SIG) Tool for Generating Stego-Images

Though there are many tools available for stego-image generation but it has following disadvantages

a) Lack of getting user desired inputs such as channels in which the changes to be incorporated, number of bits to be affected, number of rows to be affected.
b) Proper Classification of Stego-images with respect to number of rows affected
c) Proper Classification of Stego-images with respect to number of bits changed
d) Proper Classification of Stego-images with respect to number of channels changed

To overcome the above disadvantages Stego-Image Generator (SIG) for generating variety of Stego-Image has been proposed along with Steganalysis using HSI Colour Model Method.

Stego-Image Generator (SIG) Tool [23] was designed specifically for the RGB based colour Images. This SIG tool can generate 63 different kinds of Stego-Images from a single Image. The calculation is shown below

RGB channels can be chosen in seven different ways such as R, G, B, RG, RB, GB and RGB. In three different ways the Rows can be selected such as 5, 10, 20 rows. In 3 different ways the Least Significant Bits are changed such as 1, 3 or 4 bits. Reason for choosing maximum 20 rows is that in steganography message communicated will not be of much length which can be accommodated with in 20 rows. In Stego-Image maximum of 3 to 4 bits are changed in the LSB of every pixel. The complete calculation of 63 Stego-Image obtained from a single Cover-Image is shown below

Number of ways in which RGB Channels can be chosen = 7
Number of ways in which the rows can be chosen = 3
Number of ways in which the number of bits that need to changed to be chosen = 3

Total number of Stego-Images generated for single Cover Image = 7*3*3 = 63 StegoImages

---

**Algorithm Used in Stego-Image Generator**

*Input: Cover Image, Input_Channel, No of Rows to be affected (Input_Rows), No of bits to be change (Input_Bits)*

*Output: Stego-Image*
*Algorithm:*

*For 1 to Input_Row*

    *For 1 to last_col*

        *Get the Input_channel*

        *change the LSB_Bits of each pixel according*

        *to the Input_Bits*

    *End*

*End*

*Save the generated Stego_Imae with the meaningful naming convention such as*

*"filename_Inputbits_InputRows_InputChannel"*

---

The above LSB algorithm used in Stego-Image Generator covers most of the RGB Steganography algorithm such as Pixel Indicator High capacity Technique Algorithm, RGB Intensity Image stego Algorithm etc. These 63 RGB Stego-Images generated by SIG forms the superset therefore any RGB Steganalysis developed can extensively test with SIG to test their robustness of the algorithm. So far 100 Cover-Images from different categories have been given as input to SIG. Images from different categories have been chosen to have various colour combinations in the Cover-Image. Some of the sample inputs are shown in table 3.

Table 3: Input Image Category for SIG Tool

| S.No | Image Name | Category |
|---|---|---|
| 1 | Lotus.bmp | Flora |
| 2 | Monkey | Fauna |
| 3 | Baby | People |
| 4 | Sea | Natural |
| 5 | Cupcakes | Eatables |
| 6 | Tajmahal | Building |

Using the SIG tool 6300 Stego-Images have been generated. These images have been used as input in our proposed Steganalysis method using Colour model conversion.

## 5.2 Steganalysis using HSI Colour Model

There are three different types of colour models they are HSI, HSV, and RGB. Any colour model can be converted to other model using mathematical expression [21]. In HSI model the values of Signal & Image Processing : An International Journal (SIPIJ) Vol.2, No.4, December 2011Hue, Saturation and Intensity values are derived from all the three R, G and B values. Any change in the values of red or green or blue are easily reflected in all values of HSI colour model. RGB to HSI colour model can be mathematically derived with respect to normalized values of RGB and the mathematical expression is shown below

$$H = \cos^{-1}\left\{\frac{\frac{1}{2}[(R-G)+(R-B)]}{[(R-G)^2+(R-B)(G-B)]^{1/2}}\right\}$$

$$S = 1 - \frac{3}{(R+G+B)}[\min(R,G,B)]$$

$$I = \tfrac{1}{3}(R+G+B)$$

As said by Rafael C.Gonzalez et.al in the book Digital Image Processing [22] any small changes done in the RGB colour image is reflected in all components of HSI this concept has been used by the proposed method. Given any image as input in our proposed model it will be converted in to HSI Colour Model and by careful observation stego-image can be differentiated from the Cover Image. The Major Steps involved in Steganalysis using HSI Colour Model is given below

i. Given image is converted in to HSI Colour Model Image.
ii. Perceive the image for any abrupt changes in the colour.
iii. If there is colour distortion in the HSI Colour Image then it is Stego-Image
iv. If there is no Colour distortion in the HSI Colour Image then the given image is Cover Image

## 6. EXPERIMENTAL RESULTS

The proposed Steganalysis method along with Stego-Image Generator (SIG) tool was implemented using MATLAB 7 for bmp and png images were taken as input. Though our proposed method is generic, our method was tested only for stego-images generated by LSB Steganography algorithm. Input images from various categories such as natural sceneries, birds, animals etc have chosen. Images generated from our own Stego-Image Generator (SIG) tool was given as input. There were 6300 Images in our database and all these images were given as input and the results were encouraging. The 6300 stego-images generated using Stego-Image Generator tool has been used as input in Steganalysis using Colour Model method.

Sample Screen shots of SIG Tool were shown in fig 3 and 4. The Original and Stego-Image of Ace picture is shown in RGB Colour Model in fig 5 and 6 where from visual perception it's difficult to differentiate the Cover and Stego-Image. The same figures when converted to HSI Colour Model the visual distortion is seen in the top rows of Aces Stego-Image and it shown in fig 7 and 8. Fig 9 shows some of the images represented in HSI Colour model.

207

Signal & Image Processing : An International Journal (SIPIJ) Vol.2, No.4, December 2011

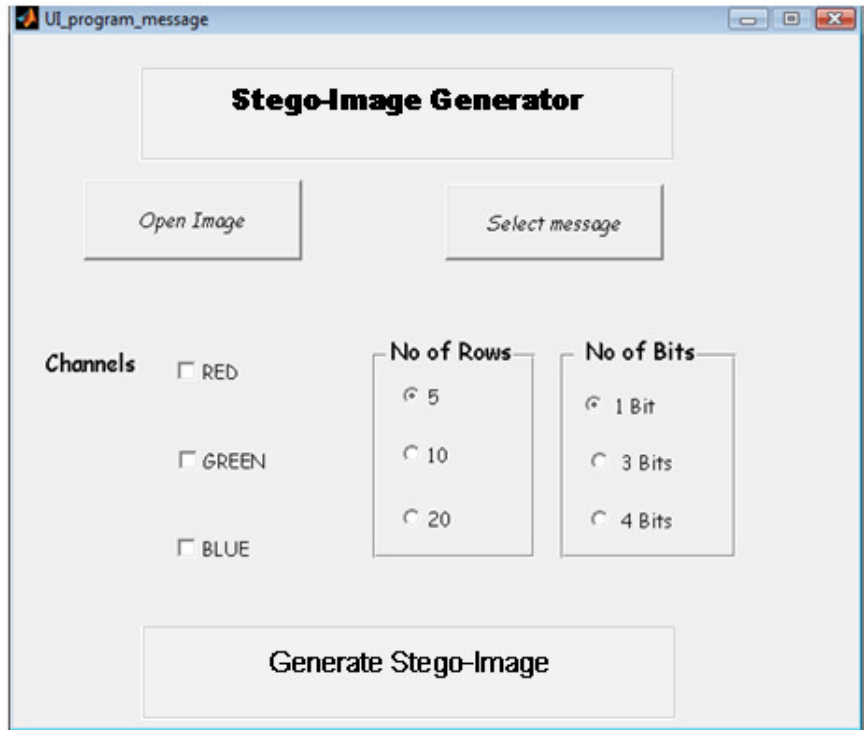

Fig 3: GUI Stego-Image Generator

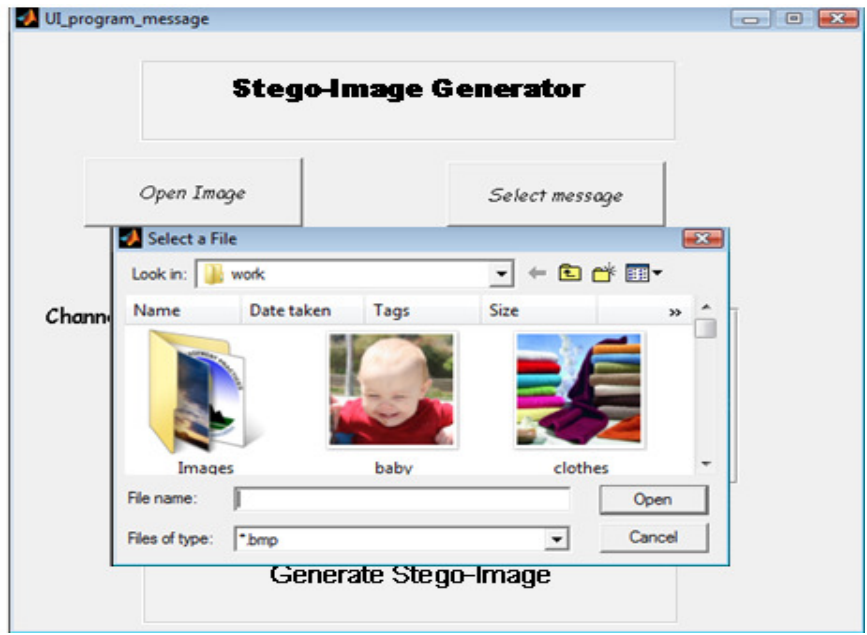

Fig 4: Selecting Cover Image for Stego-Image Generator

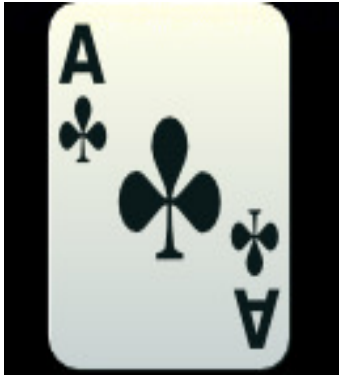 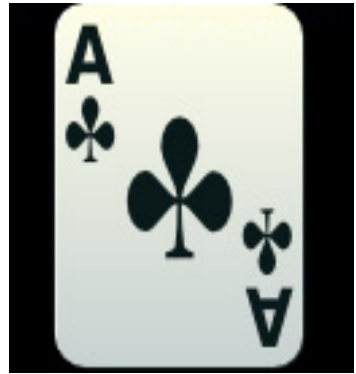

Fig 5:  Cover Image in RGB Colour Model      Fig 6:  Stego-Image in RGB Colour Model

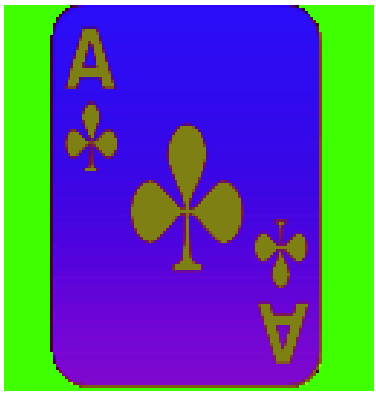 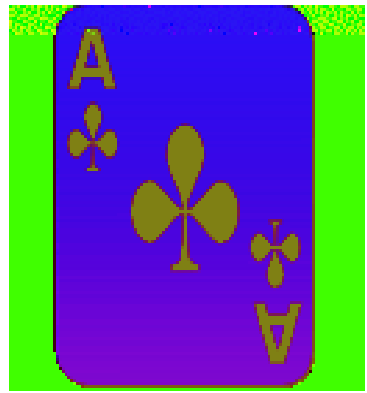

Fig 7:  Cover Image in HSI Colour Model      Fig 8:  Stego-Image in HSI Colour Model

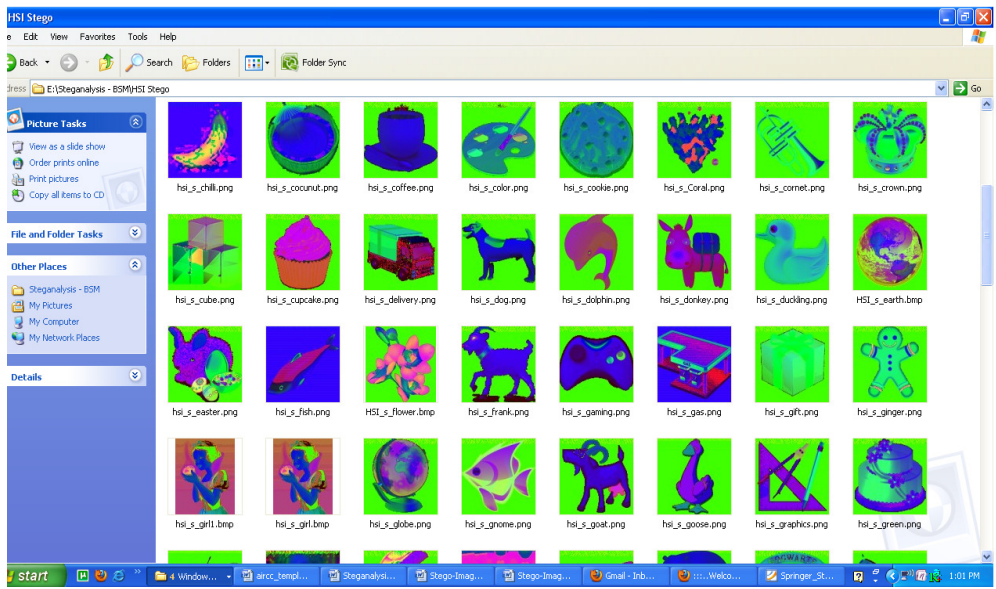

Fig 9: Sample Images in HSI Colour Model

## 7. CONCLUSION

In this paper, a technique using Colour Model conversion was proposed for discriminating the clean image and stego image (generated from LSB technique). The proposed Steganalysis method was tested on Stego-image produced by our Stego-Image Generator (SIG) tool. In future, the experiment has to be repeated in large image database available online and to test with other different Steganography algorithms.

## ACKNOWLEDGEMENT

We thank Collaborative Directed Basic Research (CDBR) - Smart and Secure Environment (SSE) project led by IIT Madras for providing excellent infrastructure. This work was supported in part by a grant received from NTRO Delhi. We would also like to thank Swarnambigai G for her support in implementation.

Signal & Image Processing : An International Journal (SIPIJ) Vol.2, No.4, December 2011

**Authors**

Mr.P.Thiyagarajan is full time Ph.D Scholar in Department of Computer Science, Pondicherry University. He obtained his Integrated M.Sc Computer Science with Distinction from College of Engineering Guindy, Anna University Main Campus, Chennai. Prior to joining research he has couple of years experience in Software Industry. His research interest includes Information security, Network Security and Database Management System.

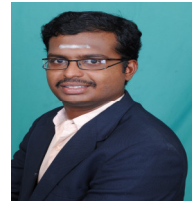

Prof. G. Aghila is currently working in the Department Computer Science, Pondicherry University. She obtained her undergraduate degree (B.E (CSE)) from TCE, Madurai in the year 1988. Her postgraduate degree M.E. (CSE) is from the School of Computer Science, Anna University, Chennai in the year 1991. She obtained her Doctorate in the field of Computer Science and Engineering from the Department of Computer Science and Engineering, Anna University, Chennai in the year 2004. She has got more than two decades of Teaching Experience and published sixty research publications in the National/ International journals and conferences. Her research interest includes Knowledge Representation and   reasoning systems, Semantic web, Information Security and Cheminformatics.

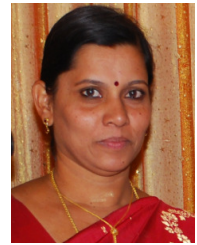

Dr. V. Prasanna Venkatesan is Associate Professor in Banking Technology Department under the School of Management at the Pondicherry University. Prior to joining the Department of Banking Technology, he was the Lecturer of Computer Science at Ramanujan School of Mathematics and Computer Science. He has published three books and more than seventy research papers in various journals, edited book volumes and conferences/workshops. He has earned B.Sc. (Physics) from Madras University and MCA, M.Tech (CSE) and PhD (CSE) from Pondicherry University, Pondicherry, India. His research interest includes Software Engineering, Object Oriented Modelling and Design, Multilingual Software Development, and Banking Technology.

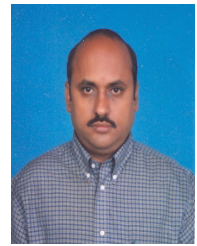